2014.03.10

# A Gaia successor with NIR Sensors

*Erik Høg and Jens Knude*

*Niels Bohr Institute, Juliane Maries Vej 30,  2100 Copenhagen Ø, Denmark*

ABSTRACT: The expected accurate astrometric data from Gaia offer the opportunity and the obligation to exploitation by a second all-sky mission. Therefore a proposal was submitted to ESA in May 2013 for a Gaia-like mission in about twenty years. Two new designs are here considered with NIR sensors in addition to the visual CCDs as in Gaia. This has been suggested by several colleagues in order to get better astrometry in obscured regions. A first design with CMOS sensors was quickly abandoned for technical reasons. The second design with specially developed CCDs for NIR looked more promising. But a deeper study showed that the longer wavelength in the near infrared at two microns deteriorates astrometry so much that only heavily obscured stars of extremely red types as M5III would obtain a better accuracy. Thus, an option with NIR sensors does not seem promising.

**Fundamental astrometry**

Design efforts on a Gaia successor should be pursued. Positive responses by email to the proposal submitted in May 2013 (Høg 2013a) and recent discussions in Copenhagen have led to the present study. We are looking much forward to comments and ideas.

The purpose of a Gaia successor in about twenty years is to perform a global scanning of the sky for astrometry as described in my proposal to ESA. The optical system of Gaia and the same scanning mode appears to be the optimal choice. The accuracy of proper motions derived from the two missions would improve by a factor of ten over Gaia.

The 1st Gaia will bring spectacular fundamental astrometry, i.e. very accurate all-sky positions, parallaxes and proper motions. A Gaia successor in twenty years would perhaps not appear as spectacular but it would bring very significant improvements and it would be unique in providing fundamental astrometry which cannot be obtained in any other way. The two Gaia missions would provide an astrometric foundation for all branches of astronomy from the solar system to compact galaxies and quasars, including radio astronomy by data which cannot be surpassed during the next 50 years.

**Six science cases for a Gaia successor**

Two missions with 20 year interval will give proper motions for a billion stars with 10 times smaller errors than from Gaia alone and the impact of this improvement on the study of Local Group galaxies is discussed by Ibata (2013). Thousands of heavy exoplanets with periods up to about 40 years can be discovered by means of data from the two missions which is not possible by any other method, Høg (2013b).

Two new designs of the payload are studied here including NIR sensors as suggested by several colleagues. The result is not encouraging since the expected gain for astrometry in obscured regions is lost because the longer wavelength at two microns deteriorates the astrometry. The proposed NIR science case can thus be abandoned.

Three further science cases are introduced by Høg (2014). The fourth case concerns discovery and mapping of optical 2D-structure in radio sources and the use of the E-ELT with 42 m aperture for the purpose is discussed. The fifth science case concerns a reference system for astrometry with the E-ELT which can lead to proper motions in clusters or dense areas with a precision of 2-6 µas/yr corresponding to about 1-3 km/s at 100 kpc distance. The sixth science case concerns the detection of QSOs solely from zero proper motion and parallax, unbiased by any assumptions on spectra. This issue is being studied by Fynbo & Høg (2014: in preparation) to see how well this can be done with the more accurate proper motions from two missions.

**Design of a Gaia successor**

Two new designs have been considered with NIR sensors in addition to the visual CCDs as in Gaia. This has been suggested by several colleagues in order to get better astrometry in obscured regions. A first design with CMOS sensors was quickly abandoned for technical reasons, see Appendix A.

The second design to be described in the following employed specially developed CCDs for NIR. This design looked more promising but a deeper study showed that the longer wavelength in the near infrared at two microns deteriorates astrometry so much that only heavily obscured stars of extreme red types as M5III would obtain a better accuracy.

CCDs are used in Gaia and have the advantage over CMOS sensors of being well suited for Time Delayed Integration (TDI). The continuous revolving scanning known from Hipparcos and Gaia could be adopted. But the quantum efficiency of CCDs based on silicon drops very quickly beyond 1100nm. A better NIR response for TDI requires development of sensors and three possibilities of development are discussed in Appendix B. Any limiting wavelength between 1 and 5 micron may be chosen by the customer in at least one of the options.

**The focal plane**

Using NIR sensors for astrometry will give a lower accuracy per detected photon as a direct consequence of the longer wavelength. The focal plane should therefore include a large number of normal CCDs as in Gaia in order to obtain the highest possible astrometric accuracy for the majority of stars, supplemented with a number of NIR sensors. Such an implementation is shown in Figure 1. The middle part with visual CCDs is identical to Gaia with respect to sky mappers and the astrometric field. The low-dispersion spectroscopy in Gaia is replaced by filter photometry with three bands in the visual and three in the NIR. This is probably sufficient to make the correction for astrometric chromaticity at every field transit. Filter photometry gives a better resolution of double stars and other extended objects. Medium-dispersion spectrometry as in Gaia is not included here.

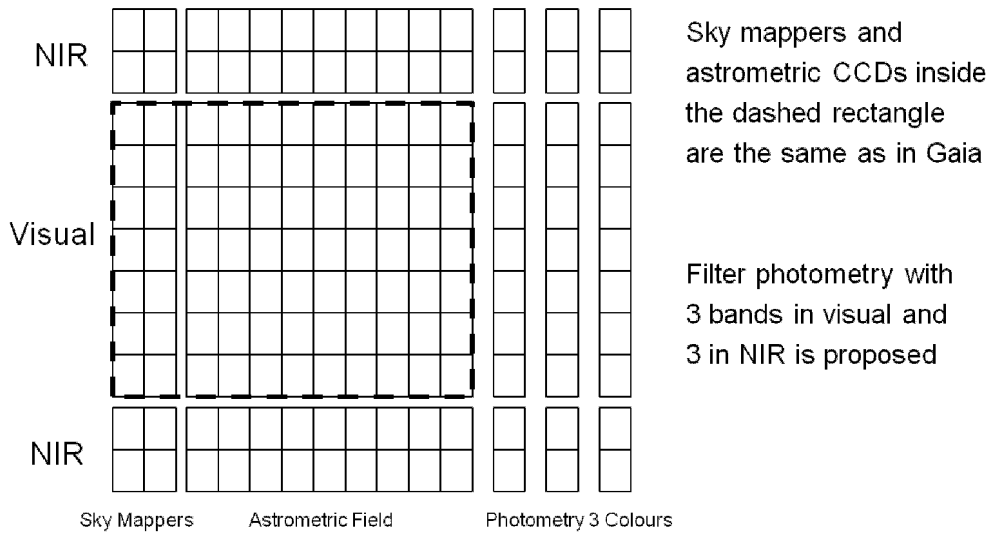

**Figure 1.** The focal plane with visual and NIR sensors.

The optical aberrations are larger in the field above and below the visual CCDs, but due to the longer wavelength of NIR the images might still be nearly diffraction limited. The pixels could be larger for the same reason without affecting the astrometric error. A careful analysis of the whole optical system would be required in a feasibility study, including a redesign with larger secondary and tertiary mirrors to avoid vignetting.

The size of area in Figure 1 reserved for NIR sensors is a suggestion. The astrometric error per detected photon in the NIR at 2 microns would necessarily be about 3 times larger than at visual wavelengths because the diffraction image is 3 times larger. The pixels could be 3 times larger, i.e. 30x90 $\mu m^2$ without affecting the accuracy.

**Scientific performance of NIR astrometry**

Astrometry in G-band and K-band according to an implementation as in Figure 1 is evaluated in the following section. The astrometric accuracies obtained with the NIR sensors and with the visual sensors are compared. The result is that only for stars with G>K+6.8 mag will the astrometric accuracy in the K band be better than in the G band. This means that even an extremely red star as M5III will only get better astrometry in the K-band if it is further reddened by an interstellar absorption of $A_V$>7 mag.

The $A_V$ vs. G-K diagram in Figure 2 shows a selection of main sequence and giant stars. It is based on standard tables of V-I and V-K colors and on the study of Gaia broad band colors by Jordi et al. (2010).

The PM2000 catalog by Ducourant et al. (2006) was studied. It contains approximately 2.5 million stars between the declinations 11 and 18 deg and it provides photometry in V and JHK from 2MASS. It was found that only less than 0.4% has V-K>6.8 mag. The two color H-K vs. J-H diagram showed that this sample for the greater part consists of late type giants but also includes some T Tau candidates.

We finally conclude that the proposed NIR astrometry is not worth the required effort.

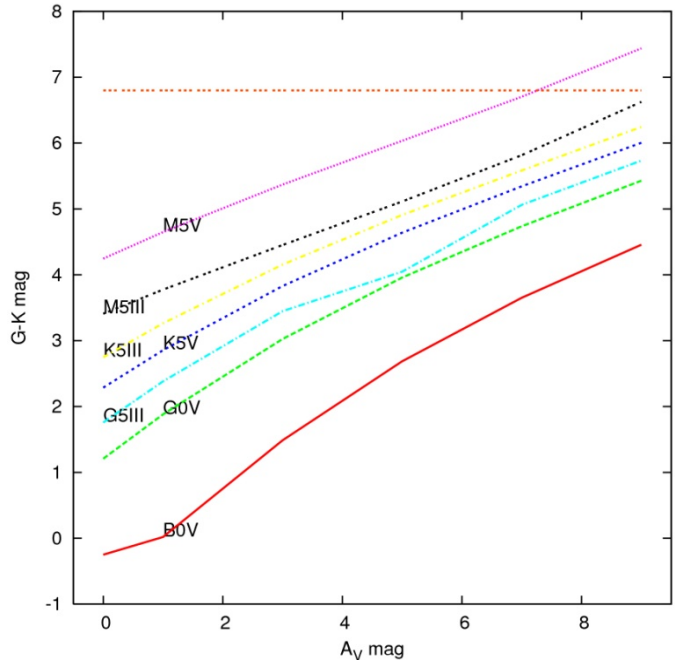

**Figure 2.** The magnitude difference G-K vs $A_V$ for spectral types from B0V to M5III.
Only if G-K>+6.8 mag (dotted line) will the astrometric accuracy in the K band be better than in the G band.

**Astrometry in G-band and K-band**

The standard Ks-band is centred on 2.2μm and has a half-value width 0.262 μm. A star of zero mag in this band gives a flux of f=121880 photons cm$^{-2}$ s$^{-1}$, according to the calibration by Cohen et al. (2003). The flux will be f=232000 in 0.5 μm width and this is the width we have agreed to assume for our NIR astrometry.

The Gaia telescope has a mirror of 1.45x0.5 = 0.725 m$^2$. Optical transmission is 0.9 as in Table 3.6 of ESA (2000). QE=0.8 as Table 3.15 at 500 nm. This gives f=1.21 10$^9$ e$^-$ s$^{-1}$ detected photons for K=0 mag.

The central wavelength is 0.6μm for G0V in Gaia. This implies that 1 electron in Gaia will give the same astrometric accuracy as $(2.2/0.6)^2$ ~14 electrons in NIR astrometry. This corresponds to 2.9 mag difference.

The stellar flux is 24.4 10$^9$ detected photons s$^{-1}$ in Gaia for G=0mag of an unreddened star of A0V. This is derived from p.239 in ESA (2000), taking into account the smaller aperture in Gaia than in GAIA.

Comparing with the flux in K for a star of K=0mag the flux ratio is 1.21/24.4=0.0496 which implies a difference of 3.26 mag.

The number of rows of CCDs in Figure1 is 4 for NIR and 7 for G which means 0.61 mag difference.

Taking this together, the astrometric accuracy obtained in Gaia at G mag will be obtained with NIR astrometry in the agreed band of 0.5 μm placed at the K band for stars with K=G-2.9-3.26-0.61=G-6.8 mag.

**Acknowledgements**

One of us (EH) is grateful for discussion and correspondence with Michael I. Andersen, Johan Fynbo, John Leif Jørgensen, Lennart Lindegren, Michael Perryman, Timo Prusti, Frederic Safa, Catherine Turon, and Sandra Vogt.

**2014.01.10** # Appendix A

## Gaia successor with CMOS

**ABSTRACT:** A payload option with CMOS sensors for a Gaia successor was considered at first but is not recommended. The purpose was to include NIR, but the better solution with NIR sensitive CCDs was soon found as described above and in Appendix B.

An option with CMOS sensors for visual and for near-IR has been considered because it seemed to be an interesting option after initial discussions with technical experts and colleagues and searches on the internet. Advantages of CMOS are less photon noise, less sensitive to radiation damage, weaker blooming effect, less power consumption and less expensive. The market for CMOS sensors is enormous and the development goes fast I learnt. But recently I learnt from a colleague that the development of CMOS does not go as fast as expected and I have returned to CCDs. Modern CCDs have nearly 100% QE while CMOS does not come above 80% and the CCD is suited for TDI.

A CMOS sensor can only be used in pointing mode because the charges cannot be shifted as in a CCD. A solution to this problem has been found but this gives others problems. Let us assume a design with telescopes and a focal assembly similar to Gaia, except that all CCDs are replaced by CMOS of the same size. Also the pixels have the same size 10x30 $\mu m^2$, elongated in the across scan direction, and a sensor array consists of 4500x1966 pixels. The satellite rotates in six hours corresponding to 60 arcsec/sec in the focal plane and the spin axis follows a revolving scanning scheme. Every star would move 264 arcsec across a CMOS in the time of *4.4 s*, resulting in a smeared image if nothing were done.

The idea is to freeze the image of the star field on the sensors for a length of time *T* by means of a tilting flat folding mirror between the pupil combiner and the focal plane. We assume here that the pupil combiner itself would be tilting with an accurate speed compensating the satellite rotation. The tilt speed must be 300"/s because the pupil magnification is 1/10 and a factor 2 is due to the reflection. For an exposure time of *T=2s* the tilt would become 600".

A tilting mirror may perhaps be implemented as an electro-optical device? The images of all sensors are read out, in parallel or sequentially, by means of an amplifier associated with every pixel in the sensor and this must be completed while the image is still frozen, resulting in sharp images along scan, but smeared somewhat across scan due to the spin axis motion.

We assume that the sky mappers are used for detection of stars brighter than a certain magnitude as in Gaia and that only data from a window around each star is transmitted to ground.

After the reading is completed the mirror must be tilted back to begin a new scan. We assume for simplicity that the reading takes negligible time compared to the integration time and that the back tilting also takes negligible time, altogether a bit more than *T* for a complete cycle. Obviously, no useful integration of light can take place during the return of the tilting mirror. At the beginning of each tilt cycle all pixels must be read and emptied lest the recorded image would be distorted.

Every star would be integrated during *T* on all sensors, also on the sensors with colour filter. But there is a limit on the allowed integration time. The images come out of focus at the beginning and end of the field during the beginning and end of freezing the image. This is caused by a tilt of the image plane due to the tilting mirror. The problem may be solved by shorter exposure times and/or a redesign of the optical system. It is concluded in the following section that an exposure time of 0.5s would give close to perfect optical resolution, requiring a total tilt of 150" of the pupil combiner. The total amplitude at the ends of the 192mm tilting mirror would be 0.07mm. A mechanical device is required to do 300 million tilts during a 5 year mission. A mechanical device would introduce a jitter which must be kept low.

It is presently concluded that the use of CMOS is not promising, but the considerations seem worth to keep in this report since further ideas and studies may be triggered.

**Tilting mirror**

The Gaia exit pupil combiner (EPC) consists of two flat mirrors M4. It is placed at a distance D in front of the focal plane to where the light arrives via two flat folding mirrors. The arrangement appears from the figure in Gaia (2011) and the dimensions are obtained from GAIA.ASF.TCN.PLM.00068, kindly supplied by Sandra. According to p.13ff, D=350 cm, the flat (curvature ~500km concave) mirrors M4 have mechanical size 192x72 mm, the exit pupil magnification is 1/10x, the beam has an aperture ratio along scan R=145/3500=1:24.The M4s are mounted separately as shown in, e.g., figure 5.4.3/11 "M4 opto-mechanical design – as manufactured. The 2 M4 are mounted on the folding optical structure through a SiC bracket." - Distribution of this description has been approved by Astrium, Francois Chassat.

The astrometric field (AF) measures b=42 cm along scan and the same across, it is covered by 9 CCDs along scan, placed in 7 rows. A CCD covers 4.73x6.0 cm^2 and is 93% filled with sensor pixels. A pixel measures 10x30 mum^2.

If the EPC tilts by an angle v the image will move d=2vD in the focal plane and the image plane will tilt by an angle 2v. The speed v' of tilting must be such that the image is frozen, i.e. the scan speed of 60"/s must be compensated which corresponds to 60/5.89mm/s in the focal plane. This requires v'=300"/s because the pupil magnification is 1/10 and the reflection gives a factor 2.

The image plane will tilt by an angle 2v so that it will be off the sensors by a=vb at the beginning and end of the field and be in focus at the field centre. The patch of light from a star on the sensor will have a length along scan of aR. This length should be smaller than the pixel, but let us set the tolerance for this maximum value to 0.010 mm.

The tilting should of course be centred on a middle position at which the image plane coincides with the sensors. This means a maximum tilt of the image plane 2v'T/2 at the ends of exposure and a maximum off-focus of 0.5bv'T at the ends of the astrometric field. The resulting maximum image smear along scan is S=0.5Rbv'T= (1/24)x0.5x420x300 T /206265 mm = 0.013T mm. An exposure time T=2s gives S=0.026 mm which is 2.6 times the value allowed above.

It follows that an exposure time of 0.5s would give S=0.006mm. A maximum smear of this amount at the ends of the astrometric field seems perfect and the smear at the sky mappers and at the photometric field would still be acceptable. The M4 should rotate around its middle with a total tilt of 150". The resulting amplitude at the ends of the 192mm mirror would be 0.07mm.

# Appendix B

# Sensor for NIR with TDI capability

**Abstract:** The scanning capability of CCDs (Time Delayed Integration, TDI) is crucial for an all sky scanning, but CCDs based on silicon are not sensitive in the NIR beyond one micron. This problem could perhaps be solved by methods proposed by two colleagues in Copenhagen.

The following is based on information from Michael I. Andersen (Technology Manager at the Niels Bohr Institute, Copenhagen University) and from John Leif Jørgensen (Professor and leader of the division for measurement and instrumentation at the Danish Technical University).

Michael Andersen mentioned three possibilities for NIR sensors with TDI capability. The most promising technique seems to be a development of the present standard technique for the hybrid HgCdTe-CMOS, used in most NIR sensors today, e.g. in the coming ESA mission Euclid. The detector material is HgCdTe bump-bonded onto the silicon readout array. Bump-bonding means: place a droplet of indium on every indiviual pixel of the silicon, then press the detector material on. The cut-off wavelength can have any value between 1 and 5 micron in the NIR as chosen by the customer. The cut-off is controlled by the fraction of cadmium, see Fig.10 in Blank et al. (2012). The technique is used for the Euclid detectors delivered by Teledyne. Such a detector for a ground based instrument cost 300 000$ for a 2kx2k sensor with 15x15 micron pixels. This technique for CMOS could probably be developed for CCD, a basic development cost of 1 M$ may be guessed for a prototype device on the TRL5 level.

Another possibility is to use germanium instead of silicon as the detector material. This could be a very expensive option because it requires development of a completely new process technique, different from that used for silicon. Germanium has a bandgab at 1.7 micron which would be the limiting wavelength. The QE might be as high as for silicon. More cooling would be required, perhaps -80C instead of -50C for silicon.

Finally, the up-converter technique could be considered, but it is very complicated because it requires pumping with laser light via a special optical system where this light is mixed with the star light.

John Leif Jørgensen wrote: *"A very attractive sensor technology likely to dominate the IR range in 5-10years time, is non-linear xtal upconversion, where a laser pumped xtal is used to convert the IR signal up into the visual range such that a standard CCD or CMOS sensor can be used. Although a more complex setup is required, this technology trivially follows the development in CCD and CMOS, and further through the optical amplification in the xtal eliminates read-out noise and other low-light noise sources in the detector."*